\documentclass[12pt,epsfig]{article}
\usepackage{graphicx}

\usepackage[latin1]{inputenc}
\usepackage{amsmath}
\usepackage{amsfonts}
\usepackage{amssymb}
\usepackage{graphicx}
\DeclareGraphicsExtensions{.pdf,.png,.jpg}

\usepackage{epstopdf}
\def\beq{\begin{equation}}
\def\eeq{\end{equation}}
\def\bea{\begin{eqnarray}}
\def\eea{\end{eqnarray}}

\begin{document}

\begin{center}
{\Large \bf Locally finite free space as limiting case of $PT$-symmetric medium
  }

\vspace{1.3cm}

{\sf   Mohammad Hasan  \footnote{e-mail address: \ \ mhasan@isro.gov.in, \ \ mohammadhasan786@gmail.com}$^{,3}$,
Mohammad Umar \footnote{e-mail address: pha212475@iitd.ac.in}
Bhabani Prasad Mandal \footnote{e-mail address:
\ \ bhabani.mandal@gmail.com, \ \ bhabani@bhu.ac.in  }}

\bigskip

{\em $^{1}$Indian Space Research Organisation,
Bangalore-560094, INDIA \\
$^{2}$ Indian Institute of Technology, Delhi-110016, INDIA. \\ 

$^{3}$Department of Physics,
Banaras Hindu University,
Varanasi-221005, INDIA. \\}

\bigskip
\bigskip

\noindent {\bf Abstract}

\end{center}
We explicitly prove that the transfer matrix of a finite layered $PT$-symmetric system of fix length $L$ consisting of $N$ units of the potential system `$+iV$' and `$-iV$' of equal thickness becomes a unit matrix in the limit $N \rightarrow \infty$. This result is true for waves of arbitrary wave vector $k$. This shows that in this limit, the transmission coefficient is always unity while the reflection amplitude is zero for all waves traversing this length $L$. Therefore, a free space of finite length $L$ can be represented as a $PT$-symmetric medium.

\medskip
\vspace{1in}
\newpage

\section{Introduction}        
Around two decades ago, it was discovered that certain class of non-Hermitian Hamiltonian can support real energy eigen values provided the Hamiltonian is invariant under a combine parity (P) and time-reversal (T) operation \cite{bender_pt_paper}. It was also noted that that a fully consistent quantum theory can be developed for non-Hermitian system in a modified Hilbert space through the modification of inner product which restore the equivalent Hermiticity and the unitary time evolution of the system\cite{mos, benr}. Since then a new dimension in quantum mechanics has emerged known as $PT$-symmetric quantum mechanics \cite {pt_book}. The non-Hermitian Hamiltonian display several new features which are originally absent in Hermitian Hamiltonians. The important features are exceptional points (EPs) \cite{ep1, ep2}, spectral singularity (SS) \cite{ss1}-\cite{aop2}, coherent perfect absorption (CPA) \cite{aop2}-\cite{cpa5}, critical coupling (CC) \cite{cc1}-\cite{cc4} and CPA-laser \cite{cpa_laser1}. Others notable features are invisibility  \cite{inv1, inv2, inv3} and reciprocity  \cite{resc}. CPA and SS have also been studied in the context of non-Hermitian space fractional quantum mechanics \cite{nh_sfqm}. Phenomena of SS have also been studied in the domain of quaternionic  quantum mechanics \cite{qqm}.

A quantum vacuum has fluctuations due to particle and anti-particle creations and annihilations in such a way that creations and annihilations balances each other to keep the net charge neutrality of the vacuum. In a sufficiently  small time interval, a snap shot of the vacuum will contain equal number of particle/anti-particles pairs or waves of complex frequencies having positive and negatives components of imaginary part. If the creation and annihilation of the particles are respectively `gain' and `loss' component for the vacuum system (and vice-versa for anti-particles), then the snap shot of the vacuum will be a $PT$-symmetric system. In other words quantum vacuum is stable under $PT$-symmetry. 

Our motivation for the present work arises due to the arguments presented in the above paragraph. In order to check that whether vacuum can be represented as $PT$-symmetric system, we consider a finite layered $PT$-symmetric system of fix length $L$ consisting of $N$ units of the potential systems `$+iV$' and `$-iV$' of equal thickness `$b$' without any intervening gap between the individual potential systems. It is shown that in the limit $N \rightarrow \infty$ such that $2Nb=L$, the entire $PT$-symmetric system of finite length $L$ is equivalent to an empty space of length $L$ in all aspect. We prove this by showing that the transfer matrix of our $PT$-symmetric system of length $L$ is a unit matrix in the above limiting case for particles of any wave vector $k$ incident on this system.         
    
We organize the paper as follows: In section \ref{transfer_matrix_section} we briefly discuss the transfer matrix for one dimensional scattering. In section \ref{layered_pt_section} we calculate the transfer matrix for our layered $PT$-symmetric system and evaluate the limiting case $N \rightarrow \infty$ of the finite length $PT$-symmetric medium in section \ref{special_pt_section}. We present results and associated discussion in section \ref{results_discussions}.  

\section{Transfer matrix} 
\label{transfer_matrix_section}
The Hamiltonian operator in one dimension for a non-relativistic particle is (in the unit $\hbar=1$ and $2m=1$)
\begin{equation}
H=-\frac{d^{2}}{d x^{2}}+ V(x) ,
\label{hamiltonian_operator}
\end{equation}
where $V(x) \in C$ . $V(x) \rightarrow 0$ as $x \rightarrow \pm \infty$. If $\int U (x) dx$, where $U(x)=(1+\vert x \vert) V(x)$ is finite over all $x$, then the Hamiltonian given above admits a scattering solution with the following asymptotic values
\begin{eqnarray}
\psi (k,x \rightarrow +\infty)= A_{+}(k) e^{ikx}+B_{+}(k) e^{-ikx}   \label{right_side} ,\\
\psi (k,x \rightarrow -\infty)= A_{-}(k) e^{ikx}+B_{-}(k) e^{-ikx}  \label{left_side} .
\end{eqnarray}    
The coefficients $A_{\pm}, B_{\pm}$ are connected through a $2 \times  2$ matrix $M$, called as transfer matrix as given below,
\beq
\begin{pmatrix}   A_{+}(k) \\ B_{+}(k)     \end{pmatrix}= M(k) \begin{pmatrix}   A_{-}(k) \\ B_{-}(k)    \end{pmatrix} .
\label{transfer_matrix}
\eeq
Where,
\beq
 M(k)= \begin{pmatrix}   M_{11}(k) & M_{12}(k) \\ M_{21}(k) & M_{22}(k)   \end{pmatrix}  .
\eeq
With the knowledge of the transfer matrix $M(k)$, the transmission and reflection coefficient are obtained as, 
\begin{equation}
t_{l}(k)=\frac{1}{M_{22}(k)} = t_{r}(k), \ \ r_{l}(k)=-\frac{M_{12}}{M_{22}(k)},  r_{r}(k)=\frac{M_{21}}{M_{22}(k)} . 
\label{tl_general}
\end{equation} 
The transfer matrix shows composition property. If the transfer matrix for two non-overlapping finite scattering regions $V_{1}$ and $V_{2}$ , where $V_{1}$ is to the left of $V_{2}$, are $M_{1}$ and $M_{2}$ respectively, then the net transfer matrix $M_{net}$ of the whole system ($V_{1}$ and $V_{2}$) is
\beq
M_{net}=M_{2}. M_{1} \ .
\eeq 
The composition result can be generalized for arbitrary numbers of non-overlapping finite scattering regions. Knowing the transfer matrix, one easily compute the scattering coefficients by using Eq. \ref{tl_general}. From Eq. \ref{tl_general}, it is also seen that if the diagonal elements are unity and off-diagonal elements are zero, then we always have $t_{l}(k)=1=t_{r}(k)$ and $r_{l}(k)=0=r_{r}(k)$. This case of transfer matrix represent empty space.  
\section{Transfer matrix of layered $PT$-symmetric system }
\label{layered_pt_section}
\begin{figure}
\begin{center}
\includegraphics[scale=0.5]{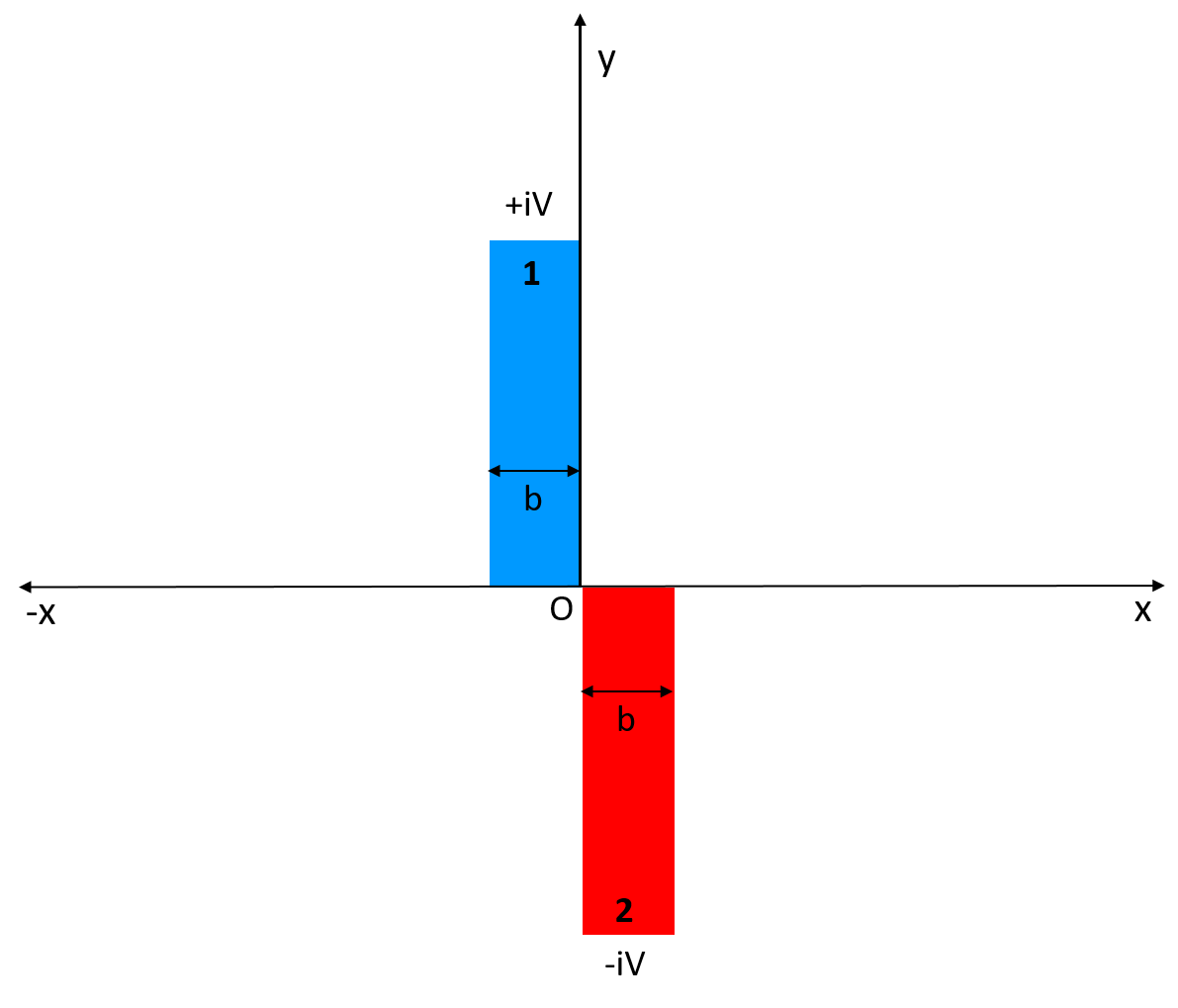}  
\caption{\it A $PT$-symmetric `unit cell' consisting of a pair of complex conjugate barrier. $y$-axis represent the imaginary  height of the potential.}  
\label{pt_barrier}
\end{center}
\end{figure}  
\begin{figure}
\begin{center}
\includegraphics[scale=0.5]{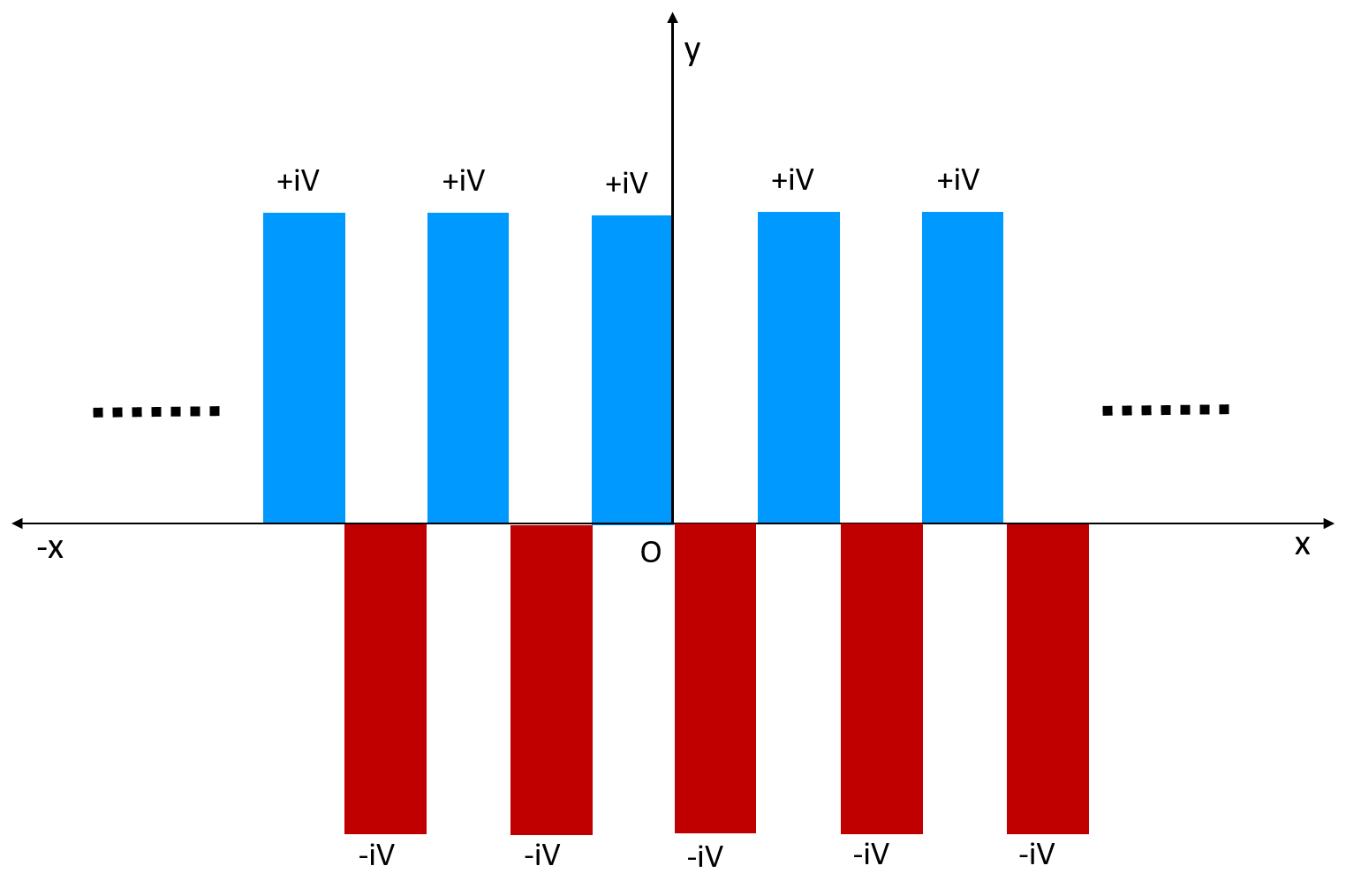}
\caption{\it A periodic $PT$-symmetric potential made by the periodic repetition of the `unit cell' potential shown in Fig \ref{pt_barrier}. $y$-axis represent the imaginary height of the potential.}  
\label{periodic_pt_barrier}
\end{center}
\end{figure}  
Fig \ref{periodic_pt_barrier} shows the layered $PT$-symmetric system which is made by periodic repetitions of `unit cell' $PT$-symmetric system shown in Fig \ref{pt_barrier}. It can be shown that the transfer matrix of the `unit cell' system is
\beq
 M(k)= \begin{pmatrix}  (\xi +i \chi ) e^{-2ikb} & i(\eta - \tau ) e^{-2ikb} \\ i(\eta + \tau ) e^{2ikb} & (\xi -i \chi ) e^{2ikb}   \end{pmatrix} . 
\label{tm_unitcell}
\eeq 
\beq
\xi= \frac{1}{2}(\cos{2\alpha}+\cosh{2\beta}) -\cos{2\phi} (\cosh^{2}{\beta} \sin^{2}{\alpha}+ \cos^{2}{\alpha} \sinh^{2}{\beta}) ,
\label{a_eq}
\eeq
\beq
\chi= \frac{1}{2} (U_{+} \cos{\phi} \sin{2\alpha} + U_{-} \sin{\phi} \sinh{2\beta}) .
\label{b_eq}
\eeq
\beq
\eta= \frac{1}{2} (\cosh{2\beta}-\cos{2 \alpha}) \sin{2 \phi}.
\label{c_eq}
\eeq
\beq
\tau= \frac{1}{2} (U_{+} \sin{\phi} \sinh{2\beta}  + U_{-} \cos{\phi} \sin{2\alpha}) .
\label{d_eq}
\eeq
In the above equations, $U_{\pm}= \frac{k}{\rho} \pm \frac{\rho}{k}$, $\alpha=b \rho \cos{\phi}$, $\beta=b \rho \sin{\phi}$. $\rho$ and $\phi$ are the modulus and phase of $k_{2}=\sqrt{k^{2}+iV}= \rho e^{i \phi}$ respectively such that $\rho=(k^{4}+V^{2})^{\frac{1}{4}}$ and $\phi= \frac{1}{2}\tan^{-1} \left( \frac{V}{k^{2}} \right )$. It can be noted that $k_{1}= \rho e^{-i \phi}$. From the knowledge of transfer matrix of a `unit cell' potential, one can find the transfer matrix for the corresponding locally periodic potential consisting N such cells \cite{griffith_periodic}. Using the approach outlined in \cite{griffith_periodic, hartman_pt}, we obtain the following transfer matrix for the layered $PT$-symmetric system,  
\beq
 \Omega (k)= \begin{pmatrix}  [T_{N}(\xi) +i \chi U_{N-1}( \xi ) ] e^{-ikL}  & i(\eta - \tau ) U_{N-1}(\xi) e^{-ikL} \\ i(\eta + \tau ) U_{N-1}(\xi) e^{ikL} & [ T_{N}(\xi) - i \chi U_{N-1}(\xi ) ] e^{ikL}   \end{pmatrix} . 
\label{layered_tm_unitcell}
\eeq 
$T_{N}(\xi)$ and  $U_{N}(\xi)$ are the Chebyshev polynomials of first and second kind respectively.  $L=2Nb$ is the net spatial extent of the layered $PT$-symmetric system.
\section{Special case of layered $PT$-symmetric medium}
\label{special_pt_section}
In this section we show that a finite length $L$ of our layered $PT$-symmetric system consisting infinitely many cells is analogous to an empty one dimensional space of length $L$. To show this we take limiting case $N \rightarrow \infty$ of each elements of transfer matrix \ref{layered_tm_unitcell} such that $b=\frac{L}{2N}$ where $L$ is fixed (and is finite). Various steps of the calculations are discussed below.

\begin{figure}
\begin{center}
\includegraphics[scale=0.65]{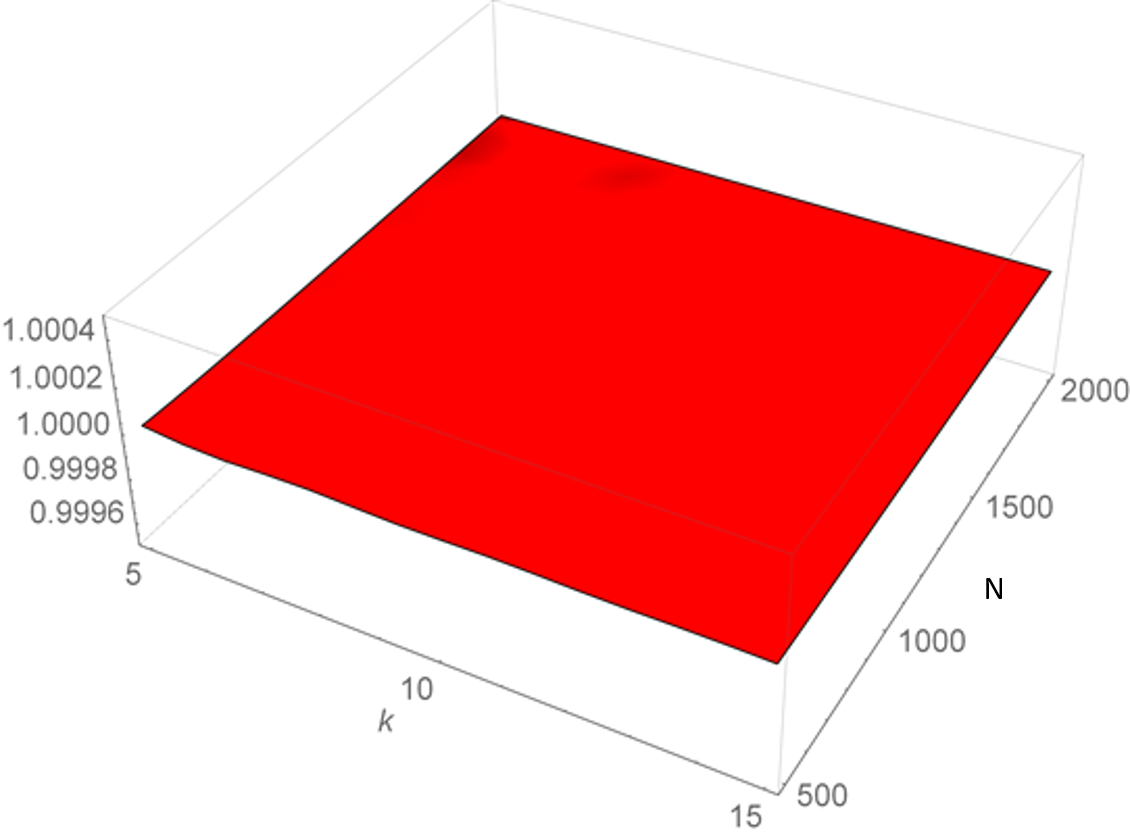}
\caption{\it Plot of transmission amplitude $T(N,k)$ as a function of $N$ and $k$ for $V=40$ and $L=1$. It is observed that the transmission is unity for large $N$. For a better clarity, the plot range of $T$ is chosen in the range from  0.9995 to 1.0005. }  
\label{tnb_figure}
\end{center}
\end{figure}

 The limiting of case of $\xi$ and $\chi$ in the leading order of $L$ can be shown to be,
\beq
\lim_{N \to \infty} \xi =1- \frac{(kL)^{2}}{2 N^{2}}, \ \ \lim_{N \to \infty} \chi = \frac{kL}{N} 
\label{limit_xi_chi}
\eeq 
In arriving at the above limit, we have used $b=\frac{L}{2N}$. We also observe $\lim_{N \to \infty} \xi < 1$. With the above limiting value of $\xi$, we also evaluate 
\beq
\lim_{N \to \infty} \cos^{-1} \xi =  \frac{kL}{N}. 
\label{limit_arccos}
\eeq 
It is further known that for $ \vert \varepsilon \vert < 1$, one can express $T_{N}{(\varepsilon)} = \cos{(N \cos^{-1} \varepsilon)}$. Therefore,
\beq
\lim_{N \to \infty} T_{N}(\xi) = \cos{ (N \cos^{-1} \xi)}. \nonumber
\eeq 
Using Eq. \ref{limit_arccos} in the above, we find
\beq
\lim_{N \to \infty} T_{N}(\xi) = \cos{kL}.
\label{tnxi_limit} 
\eeq 
Next we evaluate the limiting value of Chebyshev polynomial of second kind for the present problem. We use the following identity,
\beq
U_{N-1}(\xi) = \frac{\sin{(N\cos^{-1} \xi )}}{\sin{(\cos^{-1} \xi)}}. \nonumber
\eeq 
Using Eq. \ref{limit_arccos} in the above, the limiting value is given by,
\beq
\lim_{N \to \infty} U_{N-1}(\xi) = \frac{\sin {kL}}{\sin{ \frac{kL}{N}}}. 
\label{un_limit}
\eeq 
From the above result, we arrive at
\beq
\lim_{N \to \infty} \chi  U_{N-1}(\xi) =  \sin{kL}. 
\label{xiu_limit}
\eeq 
In the above we have used $\lim_{N \to \infty} \sin{ \frac{kL}{N}} = \frac{kL}{N} $Combining the limiting values of Eq. \ref{tnxi_limit} and Eq. \ref{xiu_limit} we have the following results,
\beq
\lim_{N \to \infty} T_{N}(\xi) \pm i \chi  U_{N-1}(\xi) =  e^{\pm i kL}. 
\label{off_limit}
\eeq 
Using Eq. \ref{off_limit}  in the diagonal values of transfer matrix (Eq. \ref{layered_tm_unitcell}) we obtain
\beq
\lim_{N \to \infty} \Omega_{11}  = 1 = \lim_{N \to \infty} \Omega_{22}, 
\label{diagonal_limit}
\eeq 
i.e. the diagonal elements are unity in the limit $N \rightarrow \infty$ provided the support $L$ is finite. Now we evaluate the limiting values of off-diagonal terms of the transfer matrix. The limiting value of $\eta \pm \tau$ in the leading order of $b$ is,
\beq
\lim_{N \to \infty} (\eta \pm \tau) = V b^{2}.
\eeq
Using Eq.\ref{un_limit} it can be easily shown that,
\beq
\lim_{N \to \infty} (\eta \pm \tau) U_{N-1}(\xi) =  \frac{V b}{2k} \sin{kL} =0.
\eeq
Therefore, the off-diagonal terms of the transfer matrix (Eq. \ref{layered_tm_unitcell}) are zero, i.e.,  
\beq
\lim_{N \to \infty} \Omega_{12}  = 0 = \lim_{N \to \infty} \Omega_{21}, 
\label{diagonal_limit}
\eeq 
for finite support $L$. Thus it is proved that for all $k$,
\beq
 \lim_{N \to \infty} \begin{pmatrix}  [T_{N}(\xi) +i \chi U_{N-1}( \xi ) ] e^{-ikL}  & i(\eta - \tau ) U_{N-1}(\xi) e^{-ikL} \\ i(\eta + \tau ) U_{N-1}(\xi) e^{ikL} & [ T_{N}(\xi) - i \chi U_{N-1}(\xi ) ] e^{ikL}   \end{pmatrix} = \begin{pmatrix}  1  & 0 \\ 0 & 1    \end{pmatrix} , 
\label{layered_tm_unitcell}
\eeq 
and therefore transmission coefficient is always unity and reflection coefficient is always zero for all wave number $k$ in the limit ${N \to \infty}$. Fig \ref{tnb_figure} shows the plot of transmission amplitude $T(N,k)$ as a function of $N$ and $k$ for $V=40$ and $L=1$. It is seen from the figure that transmission amplitude is unity for large $N$ as proven theoretically. The range of $N$ in the figure is taken from $500$ to $2000$.     

This is to be noted that when $PT$-symmetry is not respected, the vacuum configuration is not obtained. If one take the general case where the potential $+iV$ is replaced by $V_{1}+ iV_{2}$ and potential $-iV$ with $V_{1}- i \varepsilon V_{2}$, $\{ V_{1}, V_{2} \} \in R$, the net configuration of length $L$ in the limit $N \rightarrow \infty$ corresponds to a barrier potential of height $V_{1} + i(1-\varepsilon ) V_{2}$ and length $L$. When $PT$-symmetry is respected ($\varepsilon =1$), the limiting case $N \rightarrow \infty$ correspond to a real barrier of height $V_{1}$ and length $L$. The case presented in this letter is the special case $V_{1}=0,  \varepsilon =1$. The calculations for the more general case is much lengthy and is planned to be reported elsewhere.     
\section{Results and Discussions}
\label{results_discussions}
We have shown that a finite layered $PT$-symmetric system of fix length $L$ consisting of $N$ units of adjacently arranged `unit cell' $PT$-symmetric system represent a free space of length $L$ in the limit $N \rightarrow \infty$ at all wave number $k$.  The `unit cell' $PT$-symmetric system is made by potential `$+iV$' and `$-iV$' ( $V \in R^{+}$ ), of same thickness and arranged adjacently without an intervening gap.  This is proven by showing that the transfer matrix of such a layered $PT$-symmetric system over fix length $L$ is a unity matrix at all wave number $k$ for large number of unit cells. Therefore for such a system in this limit, the transmission coefficient is always unity while the reflection coefficient is always zero. Thus a free space of finite length $L$ can be represented as $PT$-symmetric medium. The result is also shown numerically for transmission coefficient.    

It is to be noted that in the present case of the layered $PT$-symmetric system, the effect of gain ($+iV$) and loss part ($-iV$) cancel each other in the limit $b \rightarrow 0$ (or $N \rightarrow \infty$ for the present problem) as the wave traverses through it. This case is different than considering vanishing strength of balanced gain and loss component of the non-Hermitian potential. The present results shows that vacuum can be represented as the special case of $PT$-symmetric medium. More investigations are needed to understand the significant of this result. This finite $PT$-symmetric system for large $L$ is also invisible for left and right incidence. If particle production is represented as $+iV$ (the gain part) and particle annihilation is represented as $-iV$ (the lossy part), then the present limiting case of layered $PT$-symmetric medium represent the static snap shot of vacuum fluctuation. However it will be worth investigating the nature of transfer matrix when the height of each `unit cell' is oscillatory  in nature where frequency of oscillation is different for different cells. This may represent a more realistic picture of vacuum fluctuation when represented in non-Hermitian quantum mechanics.

{\it \bf{Acknowledgements}}: \\
MH acknowledges supports from Director-SPO and Scientific Secretary, ISRO for the encouragement of research activities. BPM acknowledges the Research Grant for Faculty under IoE Scheme (number 6031).

\end{document}